\documentclass[twocolumn,showpacs,preprintnumbers,amsmath,amssymb]{revtex4}
\usepackage{graphicx}
\usepackage{dcolumn}
\usepackage{bm}

\begin{document}

\preprint{APS/123-QED}

\title{Constraints on Extragalactic Point Source Flux from Diffuse Neutrino Limits}
\author{Andrea Silvestri}
 \email{silvestr@uci.edu}
\author{Steven W. Barwick}
 \email{barwick@hep.ps.uci.edu}
\affiliation{Department of Physics and Astronomy, University of California, Irvine, CA 92697, USA}

\date{\today}

\begin{abstract}
We constrain the maximum flux from extragalactic neutrino point
sources by using diffuse neutrino flux limits. We show that the
maximum flux from extragalactic point sources is
${ {\rm E^{2}(dN_{\nu}/dE)} \le 1.4 \times 10^{-9}}$ 
${ (L_{\nu}/2 \times 10^{43} {\rm~erg/s})^{1/3} ~{\rm GeV~cm^{-2}~s^{-1}}}$ 
from individual point sources with average neutrino
luminosity per decade, ${ L_{\nu}}$. 
It depends only slightly on
factors such as the inhomogeneous matter density distribution in the
local universe, the luminosity distribution, and the assumed spectral
index.
The derived constraints are at least one order of magnitude below the
current experimental limits from direct searches. Significant
constraints are also derived on the number density of neutrino sources
and on the total neutrino power density.
\end{abstract}


\maketitle

The origin of ultra high energy cosmic rays (UHECR), is still
unknown. Active galactic nuclei (AGN), gamma-ray bursts (GRBs), or
processes beyond the standard model have been hypothesized to be the
sources of UHECRs.
If nearby AGN are the sources of the highest energy cosmic
rays~\cite{ref_01}, and if AGN emit
neutrinos in addition to photons, protons and other charged
particles at comparable fluxes, then individual AGN may be observable
by current generation of neutrino detectors. However, only the nearest
sources would be detectable as point sources, while the contribution
of an ensemble of unresolved extragalactic sources would generate
a diffuse flux of neutrinos. There are plausible but speculative reasons
to expect a correlation between sources of cosmic rays and sources of
neutrinos.
Several models predict a diffuse neutrino flux from AGN, in particular
neutrino production has been predicted from the core of radio-quite
AGN~\cite{ref_02,ref_03}, and from AGN jets and
radio lobes~\cite{ref_04,ref_05,ref_06}.
Direct searches for diffuse~\cite{ref_07} and point flux~\cite{ref_08}
by current telescopes have set the most stringent upper limits, but
generally have not reached the sensitivity required, and the models
suggest that challenges exist even for next generation telescopes. 
One of the primary motivations for the construction of
neutrino telescopes is to search for unexpected sources with no obvious
connection to the power emitted in the electromagnetic
band.

We show in this paper that the $\nu$-flux from extragalactic point
sources can be constrained by the measured diffuse $\nu$-flux limits,
and we also use these results to constrain the neutrino intensity
predicted in models from individual sources.
The derived constraints are one order
of magnitude below current experimental limits from direct searches
for energies between TeV-PeV, and below current limits and sensitivities
of km$^{3}$ neutrino telescopes, such as IceCube, for energies between
PeV-EeV. Since, the
constraints scale with the power of 2/3 of the measured diffuse flux,
an expected factor three improvement in the diffuse flux sensitivity
for 1 year of IceCube~\cite{ref_09} data improves the constraints by
another factor two.

Point sources of neutrinos are observed when several neutrinos
originate from the same direction, and in the context of this study,
only the very nearest of an ensemble of extragalactic sources are
detectable as point sources.
The number of detectable (or resolvable) point sources,
$N_{s}$, presented in~\cite{ref_10}, is determined for a given
diffuse $\nu$-flux limit and point source sensitivity.
The $N_{s}$ calculation is based on three assumptions:
(1) the sources are extragalactic and uniformly distributed in space;
(2) the neutrino luminosity follows a power law or broken power law distribution;
(3) the sources are assumed to emit neutrinos with an $E^{-2}$ energy
spectrum. Later, we discuss the robustness of the constraint by
investigating the validity and caveats of the assumptions.

The number of resolvable sources $N_{s}$ for a distribution of
luminosities $L_{\nu}$ per decade in energy is given by:
\begin{equation}
N_{s} \simeq \frac{\sqrt{4\pi}}{3} \frac{1}{\sqrt{{\rm ln}\left(\frac{E_{max}}{E_{min}}\right)}}
\frac{H_{0}}{c}   
\frac{K_{diff}}{(C_{point})^{3/2}}
\frac{\langle L^{3/2}_{\nu}\rangle}{\langle L_{\nu}\rangle}
\frac {1}{\xi}
\label{eq1}
\end{equation}
where the parameter $\xi$ which is close to unity, depends on
cosmology and source evolution as described in~\cite{ref_10}.
\begin{figure*}[th]
\centering
\includegraphics*[width=0.55\textwidth,angle=270,clip]{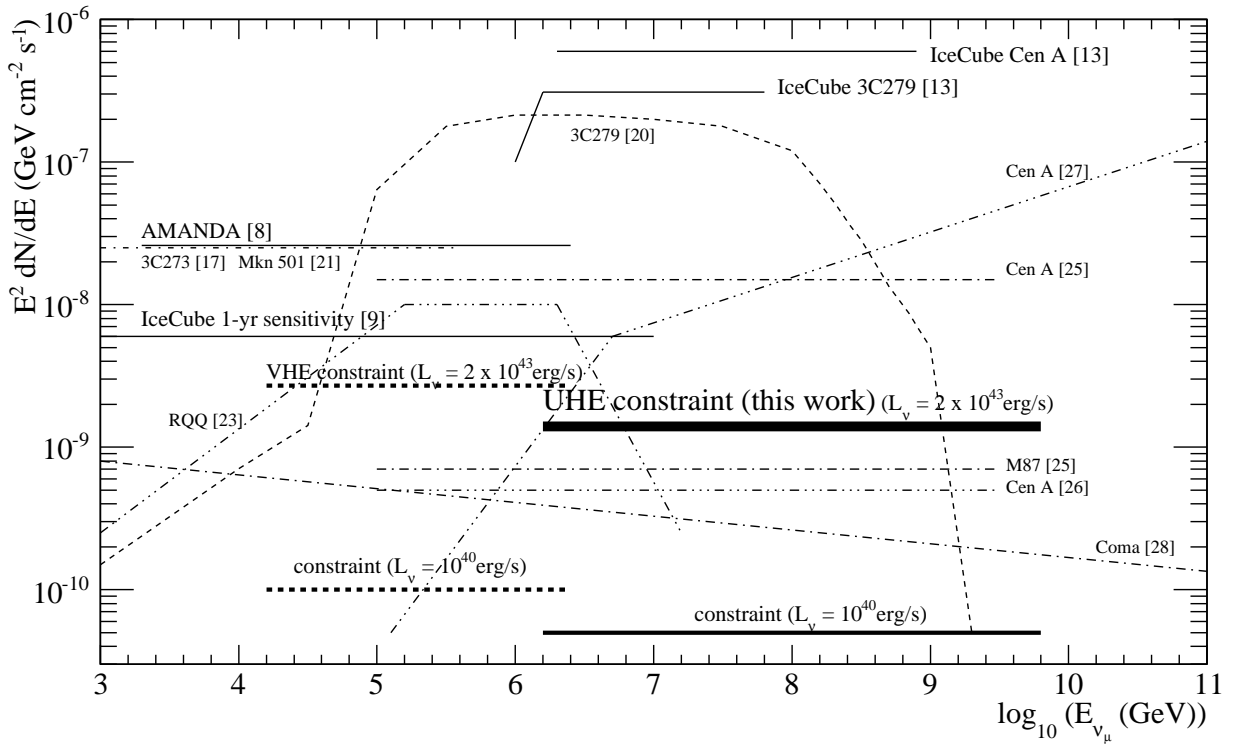}
\caption{\label{fig1}
Constraints on neutrino point fluxes derived from the UHE diffuse $\nu$-flux
limit~\cite{ref_07}, and from VHE limit~\cite{ref_11}, 
for two representative $\langle L_{\nu} \rangle= (10^{40}, 2 \times 10^{43})$
erg/s.
Current AMANDA limit~\cite{ref_08}, IceCube
sensitivity~\cite{ref_09} to neutrino point fluxes and IceCube
limits~\cite{ref_32} on fluxes from two individual point sources are
also shown (thin solid lines).
A sample of model predictions for $\nu_{\mu}$-point flux from extragalactic
sources are displayed in thin dotted-dashed lines, which are
proportional to an $E^{-2}$ spectrum or follow a broken power law.
Emission from AGN jet, calculated for a 3C279 flare of 1 day
period $[3C279]$~\cite{ref_19};
Spectra predicted for 
Mkn 501 during the outburst in 1997 $[Mkn~501]$~\cite{ref_20} and core
emission due to $pp$ interactions $[3C273]$~\cite{ref_16};
radio-quiet AGN $[RQQ]$ ~\cite{ref_22};
emission from Cen A as described in
$[Cen~A]$~\cite{ref_24},
$[Cen~A]$~\cite{ref_25} and $[Cen~A]$~\cite{ref_26};
emission from M87 $[M87]$~\cite{ref_24}, and
emission from Coma galaxy cluster $[Coma]$~\cite{ref_27}.
}
\end{figure*}
\noindent
The neutrino luminosity of the source, $L_{\nu}$ has units of (erg/s),
and $(E_{min},E_{max})$
defines the energy range of the flux sensitivity,
where $E_{max} = 10^{3}E_{min}$ for a typical experimental condition.
For canonical energy spectrum proportional to E$^{-2}$, we use the
Ultra High Energy (UHE)
results for all-flavor diffuse flux limits from AMANDA~\cite{ref_07}
to obtain the diffuse $\nu_{\mu}$-flux: $ K_{diff} \equiv 
{\rm E}^{2} \Phi_{\nu_{\mu}} =
(1/3)*{\rm E}^{2} \Phi_{\nu_{all}} =
(1/3)* 8.4 \times 10^{-8} ~~{\rm GeV}~{\rm cm}^{-2} {\rm s}^{-1} {\rm sr}^{-1}
= 2.8 \times 10^{-8} ~~{\rm GeV}~{\rm cm}^{-2} {\rm s}^{-1} {\rm sr}^{-1}$
valid for the energy interval of 1.6 PeV $<$ E $<$ 6.3 EeV.
This is the energy
interval of interest for cosmic ray interaction with energies above the ankle.
For neutrinos at the Very High Energy (VHE), we also use limits from
AMANDA~\cite{ref_11}, $K_{diff} < 7.4 \times 10^{-8}$
GeV cm$^{-2}$ s$^{-1}$ sr$^{-1}$, valid between 16 TeV to 2.5 PeV.
So, similar diffuse flux limits exist for the
entire interval from TeV to EeV energies.
$C_{point}$ is the experimental sensitivity to $\nu$-fluxes from point
sources for
an E$^{-2}$ spectrum, where the sensitivity from AMANDA~\cite{ref_08} is
$C_{point} = {\rm E^{2}(dN_{\nu}/dE)} < 2.5 \times 10^{-8}$
GeV cm$^{-2}$ s$^{-1}$.

The diffuse flux $K_{diff}$ parameter and the point flux sensitivity
$C_{point}$ are linearly correlated by the following equation:
\begin{equation}
 4 \pi K_{diff} = \left[3 ~\left(\frac{c}{H_{0}}\right) \frac{1}{r_{max}}~ N_{s}\right] \times C_{point}
\label{eq2}
\end{equation}
where $(c/H_{0})$ represents the Hubble distance given by
$c/H_{0} = 3 \times 10^{5} ~({\rm km~s^{-1}}) / 77~({\rm
  km~s^{-1}~Mpc^{-1}}) \sim 4$ Gpc.
For the case of $N_{s} < 1$ the distance ratio $(c/H_{0})/r_{max}> 1$,
which occurs for sources well within the Hubble distance.
The parameter $r_{max}$ defines
the maximum observable distance for a point source of luminosity
$L_{\nu}$, which is given by:
\begin{equation}
 r_{max} = \left[\frac{L_{\nu}}{4 \pi~ {\rm ln}(E_{max}/E_{min})~C_{point}}\right]^{1/2}
\label{eq3}
\end{equation}
The constraint also holds for time variable sources, since it depends
only on the observed luminosity and is independent of the duration of
the variability~\cite{ref_12}. Similarly, it holds for beamed
sources, such as GRBs. However for luminosities of the order of
$10^{51}$ erg/s typical of GRB emission, a dedicated
search for GRBs leads to more restrictive limits~\cite{ref_13}.

We derive an upper limit on the maximum neutrino
power density ${\cal{P}^{C}_{\nu}}$ independently of the number
density of sources, given by:
\begin{equation}
{\cal{P}}^{C}_{\nu} \le 4\pi \frac{H_{0}}{c} {\rm ln}(\frac{E_{max}}{E_{min}})
K_{diff}
= 3.4\times 10^{45} ~\frac{\rm erg/s}{\rm Gpc^{3}} 
\label{eq8}
\end{equation}
which is one order of magnitude below the power required to generate
the energy density of the observed extragalactic cosmic
rays~\cite{ref_14}.

\begin{figure*}[th]
\centering
\includegraphics*[width=0.55\textwidth,angle=270,clip]{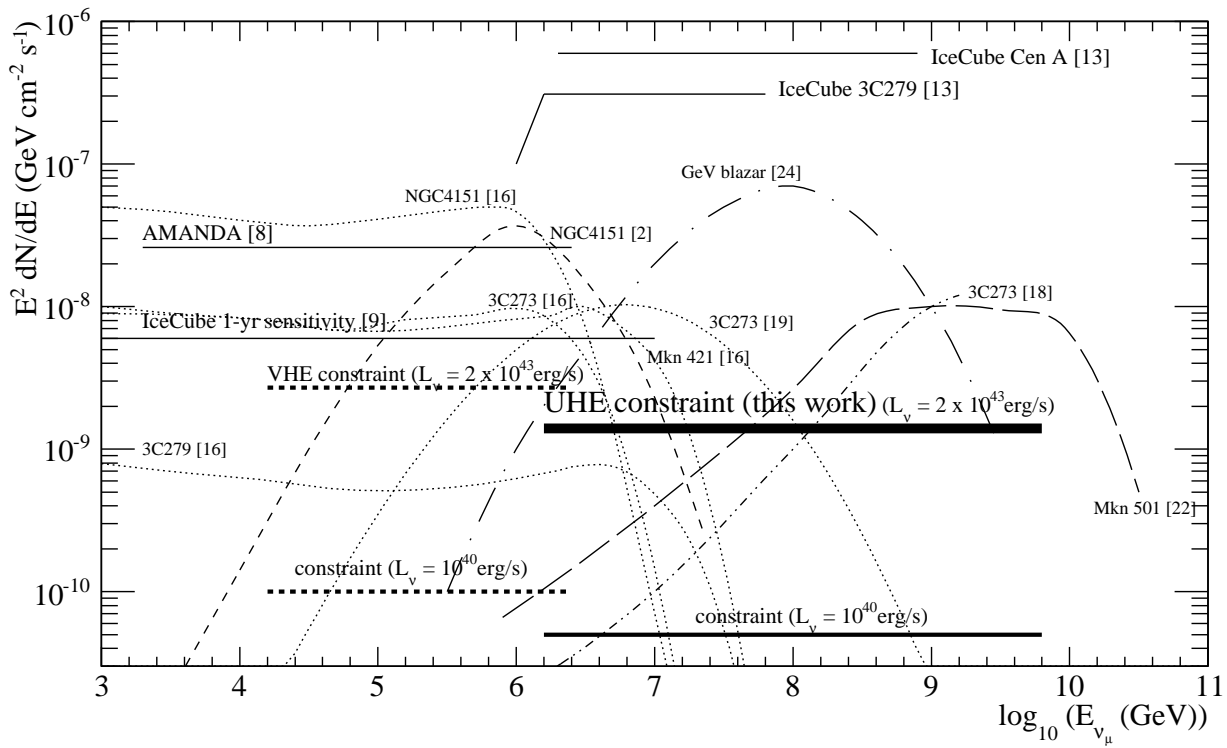}
\caption{\label{fig2}
Same flux constraints as Fig.~\ref{fig1} compared to a sample of model
predictions for
$\nu_{\mu}$-point flux from extragalactic
sources are displayed in thin dotted-dashed lines, which strongly
differ from an $E^{-2}$ spectrum.
Emission from 3C273 predicted by
$[3C273]$~\cite{ref_15},
including $pp$ and $p\gamma$ interactions
$[3C273]$~\cite{ref_17};
core emission due to $p\gamma$ interaction
$[3C273]$~\cite{ref_18}; 
AGN jet continuous emission $[3C279]$~\cite{ref_15};
emission from NGC4151 by $[NGC4151]$~\cite{ref_15} and
core emission from NGC4151 due to $p\gamma$ interaction $[NGC4151]$~\cite{ref_02};
Spectra predicted for Mkn 421 $[Mkn~421]$~\cite{ref_15},
and blazar flaring Mkn 501 $[Mkn~501]$~\cite{ref_21};
GeV-loud blazars
$[GeV~blazar]$~\cite{ref_23}.
}
\end{figure*}

A numerical value for $N_{s}$ can be estimated by incorporating
the diffuse $\nu$-flux limit and the sensitivity to point sources in
Eq.~\ref{eq1}: $ N_{s} \simeq (3.7 \cdot 10^{-29}{\rm cm^{-1}})
\times (K_{diff})\times  (C_{point})^{-3/2}\times
(L_{AGN})^{1/2}\times {1}/{\xi} \simeq 10^{-3}$
computed assuming $L_{AGN} = 2 \times 10^{43}$ erg/s.
We chose to scale the value for the neutrino luminosity $L_{AGN} =
{\cal{P}^{C}_{\nu}}/n_{s} = 2 \times 10^{43}$ erg/s for a number
density of $n_{s} \sim 10^{2}~{\rm Gpc^{-3}}$ characteristic of AGN
sources.
The parameter $\xi = \xi_{AGN} \simeq 2.2$
accounts for the effects due to
cosmology and source evolution that follows AGN~\cite{ref_10}.
The estimate for $N_{s} \simeq 10^{-3}$, which is compatible with the
non-detection of any point sources.
The constraint on $\nu$-flux is determined by resolving at least one
source, i.e. by setting $N_{s} = 1$ and
inverting Eq.~\ref{eq1} to solve for $C_{point}$:

\setlength{\arraycolsep}{5pt}
\begin{eqnarray}
{\rm E^{2} \frac{dN_{\nu}}{dE}} \le
\left[ \frac{\sqrt{4 \pi}}{3} \frac{1}{\sqrt{{\rm ln}\left(\frac{E_{max}}{E_{min}}\right)}}
  \frac{H_{0}}{c} \cdot K_{diff} {\sqrt{L_{\nu}}} \cdot \frac {1}{\xi}
  \right]^{2/3}\nonumber
\end{eqnarray}
\begin{eqnarray}
{\rm E^{2} \frac{dN_{\nu}}{dE}} \le 1.4 \times 10^{-9}
\left(\frac{L_{\nu}}{2 \times 10^{43}{\rm~erg/s}}\right)^{1/3}\left({\rm\frac{GeV}{cm^{2}~s}}\right)\nonumber\\
\label{eq4}
\end{eqnarray}
\setlength{\arraycolsep}{5pt}

\noindent
valid for the same energy range 1.6 PeV $<$ E $<$ 6.3 EeV of the
diffuse flux limit $K_{diff}$.
This result defines a benchmark flux constraint 
$\Phi_{C} \equiv {\rm E^{2}(dN_{\nu}/dE)} \le 1.4 \times 10^{-9}~{\rm
  GeV~cm^{-2}~s^{-1}}$ on neutrino fluxes from individual
extragalactic point sources that produce the power required to
generate the neutrino
flux with $L_{\nu} = 2 \times 10^{43} {\rm~erg/s}$.
The benchmark flux constraint $\Phi_{C}$
is one order of magnitude lower than present experimental limits from
direct searches, and strengthen for ensemble of sources that generate
less power. These results show that the likelihood of detecting
neutrino signal from AGN sources will be a challenge for
next generation km-scale neutrino telescopes.

Fig.~\ref{fig1} shows the benchmark constraint on extragalactic point
source fluxes derived from the UHE and VHE diffuse flux limits.
Models are shown with an energy spectrum proportional to E$^{-2}$ (or
approximatelly proportional over the UHE and VHE energy interval).
The model predictions can be compared to the derived benchmark
constraint,
$\Phi_{C}$, by assuming that the specific prediction characterizes the
mean flux $\Phi^{model}_{\nu}$, and energy distribution from an
ensemble of sources. By computing the ratio ${\cal{R}} =
\Phi_{C}/\Phi^{model}_{\nu}$, models are constrained if ${\cal{R}}<1$.
The results from the constraint $\Phi_{C}$ compared to a number of
models of neutrino point fluxes from extragalactic sources are summarized in
Tab.~\ref{tab1}.

\begin{table*}[t]
\caption{\label{tab1}
Summary of models for $\nu_{\mu}$ point flux from extragalactic
sources constrained by the results from this work.
The benchmark flux $\Phi_{C}$ defines the flux constraint for an
$E^{-2}$ spectrum, which is directly compared to the predicted
neutrino flux for a given model, $\Phi^{model}_{\nu}$.
The redshift of the source is from ~\cite{ref_28}, and the parameter
$d_{s}$ defines the distance of the source in Mpc
computed according to the relation $d_{s} = z\times c/H_{0}$. The
neutrino luminosity $L_{\nu}$ is computed from
$\Phi^{model}_{\nu}$ (see text for details).
Upper bounds on the number density, $n_{s}$, are given in units of
Gpc$^{-3}$.
The ratio ${\cal{R}} = \Phi_{C}/\Phi^{model}_{\nu} < 1$
determines a model constrained by this work.}
\begin{tabular}{@{}cccccccc@{}}\hline
Model & $\Phi^{model}_{\nu}$ & $n_{s}$ & redshift $z$& $d_{s}$ & $log_{10}(L_{\nu})$& ${\cal{R}}$ & Reference\\
 & (${\rm GeV/cm^{2}~s}$)&(Gpc$^{-3}$) &\cite{ref_28} & (Mpc) & &\\
\hline
$[3C273]$      & $1.0 \times 10^{-8}$  & $0.82 $               & 0.158339 &633   &45.2  &   $0.14$             & \cite{ref_15}\\
$[3C273]$      & $2.5 \times 10^{-8}$  & $0.33 $               & 0.158339 &633   &45.6  &   $0.06$             & \cite{ref_16}\\
$[3C273]$      & $1.0 \times 10^{-8}$  & $0.82 $               & 0.158339 &633   &45.2  &   $0.14$             & \cite{ref_17}\\
$[3C279]$      & $2.0 \times 10^{-7}$  & $3.6\times 10^{-3} $  & 0.536200 &2,145 &47.6  &   $7\times 10^{-3}$  & \cite{ref_19}\\
$[NGC4151]$    & $3.5 \times 10^{-8}$  & $5.3 \times 10^{2}$   & 0.003319 &13.3  &42.4  &   $0.04$             & \cite{ref_15}\\
$[Mkn~421]$    & $9.0 \times 10^{-9}$  & $25.3 $               & 0.030021 &120   &43.8  &   $0.16$             & \cite{ref_15}\\
$[Mkn~501]$    & $2.5 \times 10^{-8}$  & $7.2 $                & 0.033663 &135   &44.3  &   $0.06$             & \cite{ref_20}\\
$[Mkn~501]$    & $1.1 \times 10^{-8}$  & $16.4 $               & 0.033663 &135   &43.9  &   $0.13$             & \cite{ref_21}\\
$[RQQ]$        & $1.0 \times 10^{-8}$  & $8.2 \times 10^{2} $  & -        &20    &42.2  &   $0.14$             & \cite{ref_22}\\
$[Cen~A]$      & $1.5 \times 10^{-8}$  & $3.9 \times 10^{3} $  & 0.001825 &7.4   &41.6  &   $0.09$             & \cite{ref_24}\\
$[M87]$        & $7.0 \times 10^{-10}$ & $1.5 \times 10^{5} $  & 0.004360 &17.4  &41.0  &   $2$                & \cite{ref_24}\\
$[3C279]$      & $6.0 \times 10^{-10}$ & $1.2 $                & 0.536200 &2,145 &45.1  &   $2.3$              & \cite{ref_15}\\
$[Cen~A]$      & $5.0 \times 10^{-10}$ & $1.2 \times 10^{5} $  & 0.001825 &7.4   &40.1  &   $2.8$              & \cite{ref_25}\\
$[Coma]$       & $2.5 \times 10^{-10}$ & $1.5 \times 10^{3} $  & 0.023100 &92    &42.0  &   $5.6$              & \cite{ref_27}\\
\hline
\end{tabular}
\end{table*}

\begin{figure*}[th]
\centering
\includegraphics*[width=0.67\textwidth,angle=270,clip]{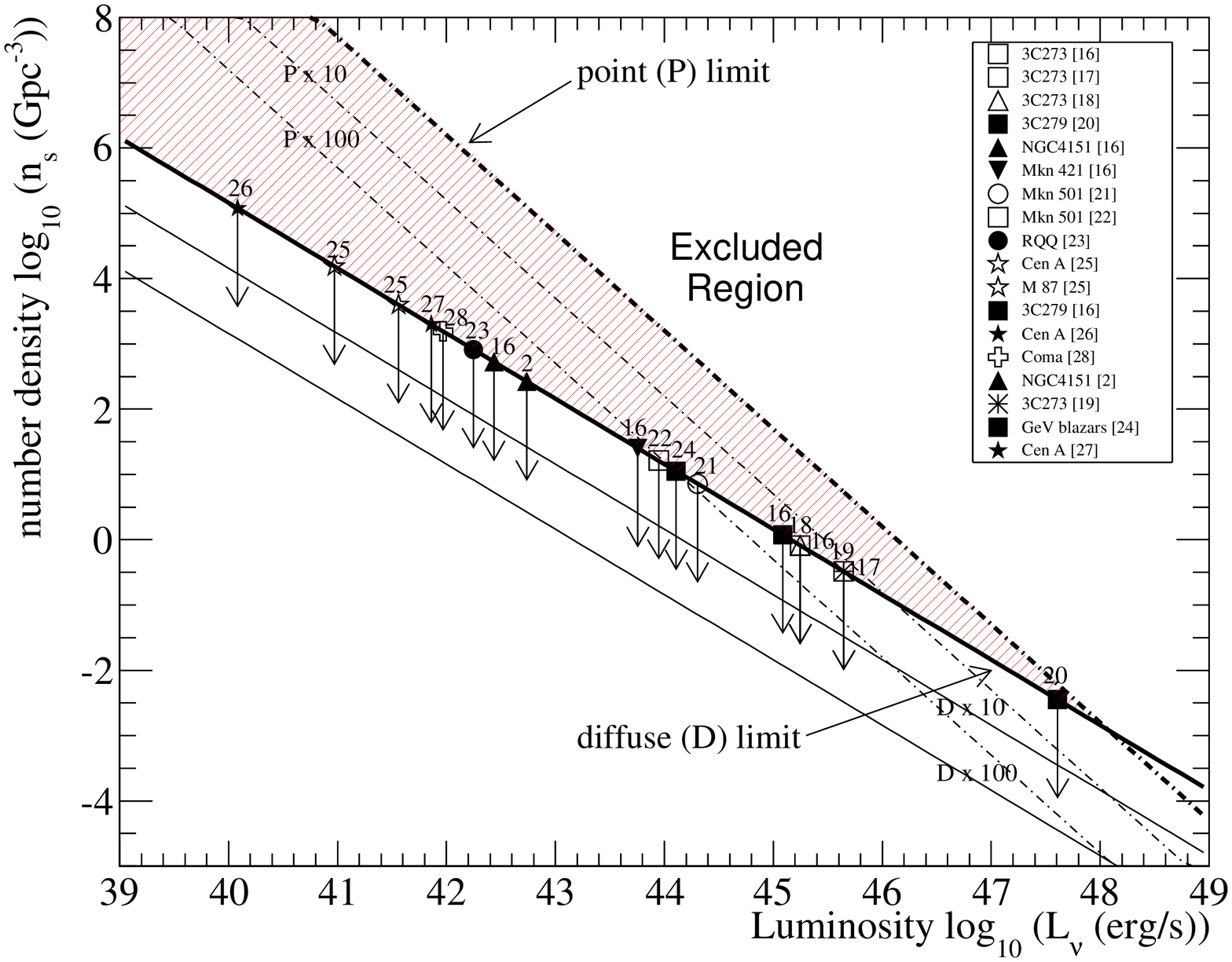}
\caption{\label{fig3}
The number density of neutrino sources $n_{s}$
plotted versus the expected neutrino luminosity predicted according to
the fluxes of the model tested. The derived upper bounds from the
diffuse flux show a stronger constraint than the limit from point
flux from direct searches. The hatched area represents the limits
accessed by the diffuse flux, but not yet accessible by direct
measurement from the point source searches. Upper bounds on number
density $n_{s}$ are computed for different neutrino sources (vertical
arrows).
Thin solid/dotted lines represent (diffuse (D)/point (P)) constraints
on the number density with one and two orders of magnitude
improvement. The region above the thick solid line is excluded by the
diffuse flux limits.
}
\end{figure*}

Models shown in Fig.~\ref{fig2} strongly deviate from an
E$^{-2}$ spectrum and in this class of models a direct comparison with
the benchmark flux $\Phi_{C}$ is less straightforward.
For the models~\cite{ref_02,ref_18,ref_23,ref_26}, the predicted energy
spectra are integrated over the UHE (VHE)
energy interval to obtain the total number
of neutrinos for the given model. The result is compared to the
integrated neutrino
events $N_{C}$ determined by the benchmark flux $\Phi_{C}$ and by the
detector neutrino effective area $A_{eff}$:
\begin{equation}
N_{C} = t_{live} \int\limits^{E_{max}}_{E_{min}} \Phi_{C} A_{eff}
(E_{\nu}) dE
\label{eq5}
\end{equation}
Similarly, the number of neutrino events expected from a given model,
$N_{model}$, is computed by substituting the predicted energy spectrum for
$\Phi_{C}$ in Eq.~\ref{eq5}. The ratio $N_{C}/N_{model}$ is found to
be 0.07, 0.2, 0.03 and 0.17 for $[NGC4151]$~\cite{ref_02},
$[3C273]$~\cite{ref_18}, $[GeV~blazar]$~\cite{ref_23}
and $[Cen~A]$~\cite{ref_26}, respectively.

The maximum number density of extragalactic sources $n_{s}$ can be
expressed in terms of the neutrino luminosity $L_{\nu}$, using the
relation~\cite{ref_12}:
\begin{equation}
n_{s} \le  4\pi \frac{H_{0}}{c} {\rm ln}(\frac{E_{max}}{E_{min}})
\times \frac{K_{diff}}{L_{\nu}}
\label{eq6}
\end{equation}
The number density is inversely proportional to the neutrino
luminosity $L_{\nu}$ and scales linearly with the measured diffuse
flux $K_{diff}$. 
Therefore we can set a constraint on the number density $n_{s}$
based on the measured diffuse flux limits $K_{diff}$, as shown in
Fig.~\ref{fig3}.
The thick solid line shows the constraint on $n_{s} \propto
K_{diff}/L_{\nu}$, so stronger diffuse flux limits constrain the
neutrino source density $n_{s}$ to lower values.
The thin parallel lines beneath it correspond to improvements in the
experimental diffuse limit $K_{diff}$ by factor of 10 and 100,
respectively.
The experimental sensitivity to point flux $C_{point}$ can also be
expressed in terms of the number density $n_{s}$ as follows:
\begin{equation}
n_{s} = 3 \sqrt{4\pi}~
\left({\rm ln}\frac{E_{max}}{E_{min}}\right)^{3/2} \times
(\frac{C_{p}}{L_{\nu}})^{3/2}
\label{eq7}
\end{equation}
Since the $n_{s}$ scales as $(L_{\nu})^{-3/2}$ the upper
bounds set by direct point searches (thick
dotted line) have a
steeper slope compared to the diffuse flux
constraints.
For neutrino luminosity
of bright extragalactic sources with values
$L_{\nu} < 10^{46}$
erg/s, the upper bounds on $n_{s}$ set by the diffuse flux are few
orders of magnitude below the bounds reached by direct searches.

We derive limits on the number density for specific source predictions
if $\Phi^{model}_{\nu}$ is assumed to characterize the average flux
for an ensemble of similar sources.
The limits on $n_{s}$ are shown as points in Fig.~\ref{fig3} and 
are summarized in Tab.\ref{tab1}.
The neutrino luminosity per energy decade is computed from the
source distance $d_{s}$ and the flux $\Phi^{model}_{\nu}$ using the
relation in Eq.~\ref{eq3},
$L_{\nu} = \Phi^{model}_{\nu} \times 4\pi d_{s}^{2}~
{\rm ln}(E_{max}/E_{min})$, which assumes isotropic emission.
The hatched area represents the parameter space accessed by the diffuse
flux constraints but not yet accessible by the point flux limits from
direct searches.
Therefore, diffuse flux limits can constrain the physics
mechanism of neutrino production from individual sources either to
lower number density or to smaller fraction of power output of
neutrino sources.
The derived limits on the number density $n_{s}$ of neutrino sources
depend only on neutrino information, without making specific
associations with source class based on electromagnetic measurements.
The region above the thick solid line is the excluded region by the
upper bounds on the number density derived from the diffuse limits. 

The thick dark horizontal line in Fig.~\ref{fig1} and Fig.~\ref{fig2}
indicates our
primary constraint $\Phi_{C}$.
We address the robustness of the constraint by focusing the discussion
on the three assumptions involved in the calculation of $N_{s}$.

The matter distribution within 5 Mpc of the Milky Way is far from 
uniform, which suggests the possibility that the local number density
of neutrino sources, $n_{l}$, may be higher than the universal
average of number density, $\langle n_{s}\rangle$. We argue that, in
practice, the local inhomogeneity affects only the class of sources
characterised by low luminosities. The bright sources are too rare to
be affected by local matter density variation - the likelihood of
finding a bright neutrino source within 5 Mpc is small to begin with
(if electromagnetic luminosity and neutrino luminosity are comparable), and the
local enhancements in matter density insufficient to change the
probability of detection.

On the other hand, if sources have a low mean luminosity, then the
nearest in the ensemble are more likely to be within a distance that
could be affected by fluctuations in the local matter density. For
example, within 4 Mpc, the ratio between local matter density to the
universal average known as overdensity is estimated to be about
5.3~\cite{ref_29}.
In this case, the flux constraint (Eq.~\ref{eq4}) should be adjusted
to account for the higher density of local matter,
$\Phi'= \Phi*(n_{l}/\langle n_{s}\rangle)^{2/3}$.
However, as Tab.~\ref{tab2} shows, the adjusted fluxes are below
$\Phi_{C}$ for a wide range of $\langle L_{\nu}\rangle$.
For distances larger than $8$ Mpc the overdensity of galaxies is
rapidly approaching the universal mass density.
To exceed $\Phi_{C}$ a source of a given luminosity $L_{\nu}$ must be
within a distance 
$d_{l} = (4 \pi/3)^{1/3}\cdot r_{max} * (\Phi'/\Phi_{C})^{1/2}$.
Assuming that the neutrino luminosity is comparable to the maximum
luminosity in any electromagnetic band, no sources are found within a
distance $d_{l}$ that would violate $\Phi_{C}$.

\begin{table}[h]
\caption{\label{tab2}
Adjusted flux constraints $\Phi'$ to account for local enhancement of
source density.}
\centering
\begin{tabular}{cccccc}
\hline
$\langle L_{\nu}\rangle$ & $\Phi$& $n_l/\langle n_{s}\rangle$ & $(n_{l}/\langle n_{s}\rangle)^{2/3}$ &$\Phi'$ &$d_{l}$ \\
erg/s  &GeV/cm$^2{\rm s}$ & \cite{ref_29} & &GeV/cm$^2{\rm s}$ & Mpc\\
\hline
$6\times 10^{41}$ & $4\times 10^{-10}$ & 5.3         & 3   & $1.2\times 10^{-9}$ & 4 \\
$2.5\times 10^{42}$ & $7\times 10^{-10}$ & 1.3          & 1.2    & $8.4\times 10^{-10}$ & 8 \\
\hline
\end{tabular}
\end{table}

We address now the assumption that the neutrino luminosity distribution
is proportional to a (possibly broken) power law, which is observed for
several classes of sources in the electromagnetic band.
It was shown in ~\cite{ref_12} that $N_{s}$ computed from the full
distribution agrees to within few percent with a simpler calculation
using only the mean luminosity of the distribution.
The reason is that the most common luminosities in the distribution
can only be observed at relatively short distances, so source
evolution and cosmological effects are negligible. Sources with
large luminosities are too rare to contribute significantly.
On the other hand, it could be argued that the unknown luminosity 
distribution function is not well described by a (possibly broken) power 
law that typifies electromagnetic sources~\cite{ref_30}. In this
scenario, by using the limit on the maximum power density in
Eq.~\ref{eq8}, it is
possible to constrain the mean luminosity for a given source class,
if the number density is known, using the relation~\cite{ref_12}:
\begin{equation}
L^{C}_{\nu} \le 4\pi \frac{H_{0}}{c} {\rm ln}(\frac{E_{max}}{E_{min}})
\frac{K_{diff}}{n_{s}}
= \frac{{\cal{P}^{C}_{\nu}}}{n_{s}} ~{\rm erg/s} 
\label{eq9}
\end{equation}
For AGN selected in the x-ray band,
$n_{s} \sim 1.4\times 10^{4} ~{\rm Gpc^{-3}}$~\cite{ref_31}, and the
mean neutrino luminosity is $L^{C}_{\nu} < 2.4 \times 10^{41}$ erg/s,
which is approximately two orders of magnitude lower than the average
luminosity in the x-ray band.

The constraint can be extended to energy spectra that differ from 
the assumed $E^{-2}$ dependence, but the constraint applies over a 
restricted energy interval that matches the energy 
interval of the diffuse neutrino limits. Experimental
diffuse limits span two different energy regions, VHE and UHE, and either 
limit can be inserted into Eq.~\ref{eq4}. The restriction in energy range is 
required to avoid extrapolating the energy spectrum to unphysical 
values. In other words, for power law indices far from 2, the spectrum 
must cut-off at high energies for indices $\gamma < 2$, or at low
energies for indices $\gamma > 2$. Subject to this restriction, we
find that the constraint depends weakly on the assumed spectral
index. For example, the constraints improve by a factor 2 for hard
spectra $(\gamma=1)$ and weaken by roughly the same factor for soft
spectra $(\gamma=3)$~\cite{ref_12}.

To summarize, we have presented in this paper the constraint on
neutrino fluxes from extragalactic
point sources, which is
${\rm E^{2}(dN_{\nu}/dE)} \le 1.4 \times 10^{-9} ~(L_{\nu}/2\times 10^{43}
{\rm~erg/s})^{1/3}~~{\rm GeV~cm^{-2}~s^{-1}}$. These constraints 
are one order of magnitude below current experimental
limits from direct searches if the average $L_{\nu}$
distribution is comparable to the electromagnetic luminosity that characterizes the
brightest AGN.
As experimental data improves the derived constraints on fluxes from extragalactic
sources ${\rm E^{2}(dN_{\nu}/dE)} \propto K_{diff}^{2/3}$ improves
with the diffuse flux limits to the $2/3$ power, while constraints on
the number density
$n_{s} \propto K_{diff}$ and the total neutrino power density
${\cal{P}^{C}_{\nu}} \propto K_{diff}$ improve linearly with the
diffuse limits.
We tested a number of model predictions for $\nu$-point
fluxes, and models which predict fluxes
higher than the benchmark
constraint have been restricted by this analysis.
The constraint is strengthened for less luminous sources, and
noncompetitive with direct searches for
highly luminous explosive
sources, such as GRBs. We found that the constraint is robust when
accounting for the non-uniform distribution of matter that surrounds
our galaxy, or considering energy spectra that deviate from $E^{-2}$, or
various models of cosmological evolution.
We also derived an upper limit on the maximum neutrino power density
which is significantly below the observed power density from
extragalactic cosmic rays.
We showed that diffuse flux limits can strongly constrain the number
density of neutrino sources $n_{s}$. The constraints derived from the diffuse
limits for sources with luminosities $L_{\nu} < 10^{46}$ erg/s is stronger by few
orders of magnitude compared to the point flux limits from direct searches.
The parameter space accessed by the $n_{s}$ constrained from the
diffuse limits for sources within this luminosity range
is a challenge for direct point searches even for kilometer-cube
neutrino detectors.
The constraint suggests that the observation of extragalactic neutrino sources will 
be a challenge for kilometer scale detectors unless the source is much 
closer than the characteristic distance between sources, $d_{l}$, after 
accounting for local enhancement of the matter density.
Although the constraint cannot rule out the existence of a unique,
nearby extragalactic neutrino sources, we note that assuming $L_{\nu} \sim
L_{\gamma}$, we found no counterparts in
the electromagnetic band with the required luminosity and distance to violate the
constraint.

The authors acknowledge support from 
U.S. National Science Foundation-Physics Division, and the
NSF-supported TeraGrid system at the San Diego Supercomputer Center
(SDSC). We thank the referee for valuable suggestions and
criticisms.

\end{document}